\documentclass [12 pt]{article}
\usepackage[cp1251]{inputenc}
\usepackage[T2A]{fontenc}
\usepackage{amssymb,amsmath}
\usepackage[english]{babel}
\RequirePackage[dvips]{graphicx}
\begin{document}
\begin{center}
\textbf{PROPERTIES OF SYNCHRONIZATION IN THE SYSTEMS OF
NON -- IDENTICAL COUPLED VAN DER POL AND VAN DER POL -- DUFFING OSCILLATORS. BROADBAND SYNCHRONIZATION}\\
\medskip{Alexander~P.~Kuznetsov$^{1}$, Julia~P.~Roman$^{2}$}\\
\small{\it $^{1}$Institute of Radio-Engineering and Electronics of RAS, Zelenaya 38, Saratov 410019, Russia\\
$^{2}$Department of Nonlinear Processes, Saratov State University, Astrakhanskaya 83, Saratov 410026, Russia}
\end{center}

\begin{abstract}
The particular properties of dynamics are discussed for the dissipatively coupled van der Pol oscillators, non-identical in values of parameters controlling the Hopf bifurcation. Possibility of a special synchronization regime in an infinitively long band between oscillation death and quasiperiodic areas is shown for such system. Features of the bifurcation picture are discussed for different values of the control parameters and for the case of additional Duffing-type nonlinearity. Analysis of the abridged equations is presented.
\end{abstract}

\renewcommand{\thesection}{\normalsize{\arabic{section}.}}
\section{\normalsize{Introduction}}

\hspace*{\parindent}The system of two coupled van der Pol oscillators is the basic model of nonlinear dynamics demonstrating the phenomenon of mutual synchronization. It is of interest for investigators concerned with synchronization theory and its possible applications (see book [1] and refs. below). For example, the system of two coupled van der Pol oscillators was studied by D,G. Aronson at al. [2], D.S. Cohen and J.C. Neu [3, 4] in the case of dissipatively coupled oscillators, N. Minorsky [5], R.H. Rand with P.J. Holmes and T.~Chakraborty [6-8] in the case of weak inertial coupling, R.H. Rand  and D.W. Storti [9] and Pastor - Diaz at al. [10] in the case of strong inertial coupling. Similar system is used also in works of T. Pavlidis [11], M. Poliashenko [12, 13] and D.S. Cohen, J.C. Neu [4, 5] for modeling biological and chemical processes. Description of the system by means of Adler phase equation was presented in work [7]. However, nowadays interest to this problem does not decrease, since new aspects and oscillation effects are detected. In this respect we may point out, for example, recent resumptive work [14], where this system is studied within the framework of quasiharmonic approximation including possibility of combined coupling. However, most of the works on this subject does not go beyond the case of identical oscillators' parameters controlling the Andronov - Hopf bifurcation. Work [14] is an exception, but the special attention is also devoted there to the case of identical subsystems. Moreover, authors have restricted themselves to the assumption that controlling parameters are small, also they didn't take into account the anisochronism of oscillators. But it is anisochronism that results in the normal form of the Andronov\:- Hopf bifurcation and its consideration is important in respect to the generalization of results.

In this work we shall carry out the investigation of the parameter space structure and of possible oscillation regimes in the dissipatively coupled van der Pol and van der Pol - Duffing oscillators with non-identical controlling parameters. The initial system of differential equations describing the interaction between the oscillators is of the form
\begin{equation}
\begin{split}
&\frac{\displaystyle d^{2}x}{\displaystyle dt^{2}}-(\lambda_{1}-x^{2})\frac{\displaystyle dx}{\displaystyle dt}+x+\beta x^{3}+\mu(\frac{\displaystyle dx}{\displaystyle dt}-\frac{\displaystyle dy}{\displaystyle\displaystyle dt})=0,\\
&\frac{\displaystyle d^{2}y}{\displaystyle dt^{2}}-(\lambda_{2}-y^{2})\frac{\displaystyle dy}{\displaystyle dt}+(1+\delta)y+\beta y^{3}+\mu(\frac{\displaystyle dy}{\displaystyle dt}-\frac{\displaystyle dx}{\displaystyle dt})=0.\\
\end{split}
\end{equation}

Here $\lambda_{1}$ and $\lambda_{2}$ are parameters characterizing the excess above the threshold of the Andronov - Hopf bifurcation in autonomous oscillators, $\delta$ is the frequency mismatch between the autonomous second and first oscillators and $\mu$ is the coefficient of dissipative coupling. Parameter $\beta$ is related to the insertion of nonlinearity of Duffing oscillator type into the van der Pol equation. Within the framework of quasiharmonic approximation it is responsible for the nonlinearity in the phase equation and, correspondingly, for the anisochronism [1]. Strictly speaking, anisochronism is also possible in the system (1) for large enough parameters $\lambda_{1}$ and $\lambda_{2}$ values. But nevertheless we shall refer to the parameter $\beta$ as to the parameter of phase nonlinearity or anisochronism.

As we have already noticed, parameters $\lambda_{1}$ and $\lambda_{2}$ are responsible for the Andronov\:- Hopf bifurcations in subsystems. They define the sizes of limit cycles of autonomous oscillators in the phase space. If the values of $\lambda_{1}$ and $\lambda_{2}$ are small, the limit cycles are small too. In that case one can use quasiharmonic approximation. We shall choose  the values of $\lambda_{1}$ and $\lambda_{2}$ from the $[1, 2]$ interval, i.e.  $\lambda_{1}, \lambda_{2}\in[1, 2]$. Such choice gives, on the one hand, an opportunity to reveal features that are characteristic for the quasiharmonic approximation on the qualitative level at least and, on the other hand, to illustrate the phenomena intrinsic to the system (1) that appeared to be essentially richer.

\section{\normalsize{The structure of the (frequency mismatch -- coupling value) parameters plane for non-identical subsystems. The possibility of the broadband synchronization}}

\hspace*{\parindent}At first, we shall present the results of computer investigation of the system (1) parameter space. We shall use the method of dynamic regime chart construction (see, e.g., [15,~16]). Within the framework of such a method we shall mark the oscillation period of the system of coupled oscillators by means of dark color and gray color hue on the parameter plane (frequency mismatch $\delta$ -- coupling value $\mu$). White color corresponds to the chaotic or quasiharmonic motions. Cycle periods were calculated by means of the Poincare section method: this is the number of points of intersection of the phase trajectory on the attractor and the surface selected as the Poincare section. Only those crossings were taken into account that correspond to the trajectories coming to the surface from the one side.

\tolerance=500
The system under investigation is characterized by four-dimensional phase space $(x,\dot{x},y,\dot{y})$. Therefore three-dimensional hypersurface that is preset by means of some additional condition, e.g., zero velocity of the second oscillator $\dot{y}=0$ may serve as the Poincare section. In that case the number $n$ of points of intersection of trajectory and section was determined. Colors on the charts are chosen in accordance with the period $n$.

\tolerance=200
The chart of dynamic regimes obtained in such a way for identical van der Pol oscillators is given in fig.~1 on the $(\delta,\mu)$-plane for $\lambda_{1}=\lambda_{2}=1$. Subsequent characteristic areas shown on this chart are:
\begin{itemize}
\item main synchronization tongue with the frequency ratio of 1/1;
\item area of quasiperiodic regimes with the embedded system of higher synchronization tongues among which the tongue with the rotation number of 1/3 is the most cha\-racteristic;
\item	area of the "oscillation death" effect [1, 2] which corresponds to the  stability of the equilibrium state point at the origin in the presence of sufficiently strong dissipative coupling.
\end{itemize}

We will characterize mutual oscillations of oscillators by means of the rotation number $w$. By analogy with [16, 17] we find numerically the average return time $\tau_{y}$ for the chosen Poincare section $\dot{y}=0$ and the average return time $\tau_{x}$ for the section of the first oscillator defined by $x=0$. Then the rotation number is defined as $w=\tau_{y}/\tau_{x}$. The graph of the rotation number $w$ dependency on the frequency mismatch $\delta$ is shown in fig.~2~(a) for the coupling parameter value $\mu=0.5$. On this graph one can see characteristic wide "steps" that correspond to the rotation number values $w=1,1/3,1/5$, etc.

Discrepancy between the controlling parameter values of coupled van der Pol oscillators leads to the change of the (frequency mismatch -- coupling value) parameter plane structure. The chart of dynamic regimes for non-identical subsystems with $\lambda_{1}=1.25$ and $\lambda_{2}=1$ is displayed in fig.~3. Non-identity is important but it is still not very large. Phase plane portraits computed at several chosen points of the parameter plane are also shown in fig.~3 in pairs on $(x,\dot{x})$ and $(y,\dot{y})$-planes.

Comparing the charts of regimes in fig.~1 and fig.~3, one can see that they are essentially different in one respect. The boundary between the oscillation death area and the area of quasiperiodic regimes in fig.~3 is not a line but a band of finite width in coupling parameter $\lambda_{2}<\mu<\lambda_{1}$ that stretches infinitely into the area of increased coupling mismatch. Existence of synchronization in the system in the presence of arbitrarily large values of oscillators eigenfrequency mismatch one can call "broadband synchronization".

Moving inside this band towards the increase of the second oscillator's eigenfrequency one can observe synchronous regimes that differ from each other in "twistedness" of the phase trajectory on the phase plane portraits of the second oscillator presented in fig.~3. Corresponding chart areas look like "degraded" tops of synchronization tongues (it is mainly visible for the characteristic synchronization tongue $w=1/3$).

The graph of the rotation number $w$ dependence upon the frequency mismatch $\delta$ inside the area of the "broadband synchronization" is presented in fig.~2~(b) for $\lambda_{1}=1.25$, $\lambda_{2}=1$ and $\mu=1.05$. One can see considerable alterations in comparison with fig.~2~(a). Areas with irrational rotation numbers disappeared and the system of steps with $w=1,1/3$ and 1/5 became essentially more evident.

The dynamic regimes chart for the system of coupled van der Pol oscillators (1) is shown on the top piece of fig.~4 for the case of distinctly different controlling parameters of subsystems $\lambda_{1}=2$, $\lambda_{2}=1$. As an illustration of the second oscillator's behavior the examples of its phase plane portraits on the $(y,\dot{y})$-plane, Poincare sections and $y(t)$ - realizations are given on fig.~4. We should remind that the Poincare section was chosen as the hypersurface $\dot{y}=0$ in four-dimensional phase space $(x,\dot{x},y,\dot{y})$. Hence the Poincare map is automorphism  of the three-dimensional space $(x,\dot{x},y)$ as is shown on fig.~4. One can see fixed points and cycles with periods of 1, 3 and 5 which correspond to the regime types on the chart. The graph of the dependence of the corresponding rotation numbers upon the frequency mismatch is shown in fig.~2~(c) for $\lambda_{1}=2$, $\lambda_{2}=1$ and $\mu=1.5$. One can see steps with $w=1,1/3$ and 1/5 which are changing each other successively.

The dynamic regime chart for the system of coupled van der Pol oscillators is shown on fig.~5 for the case of even greater value of the controlling parameter of the first system $\lambda_{1}=3$, $\lambda_{2}=1$. In contrast to fig.~3, oscillators are essentially non-identical. As fig.~5 shows, besides the $\mu$-directionally broadening of the synchronization band the considerable broadening of domains with fractional rotation numbers inside the band takes place during the increase in difference of controlling parameters. These areas are filling gradually the most part of the band of broadband synchronization. So the remarkable feature of the problem is the preservation of fixed rotation number in essentially broad range of frequency which is several times more than the width of corresponding tongues for identical subsystems.

In case of distinctly different $\lambda_{1}$ and $\lambda_{2}$ parameters values one can provide an explanation of broadband synchronization appearance. It is based essentially on the non-identity of oscillators. Really, if $\mu$ exceeds both $\lambda_{1}$ and $\lambda_{2}$, both oscillators are behind the threshold of the "oscillation death" effect. But in the range of $\lambda_{2}<\mu<\lambda_{1}$ only the second oscillator appears to be essentially dissipative. And at the same time the first oscillator appears to be leading and excites, in fact, the second one. In this respect, different scales along the coordinate axes on the phase plane portraits for the first and the second oscillators are quite characteristic.

To discuss the mutual influence of coupled van der Pol oscillators in the described above domain of the parameters plane, we consider the system (1) with the "shut off" influence of the second oscillator upon the first one
\begin{equation}
\begin{split}
&\frac{\displaystyle d^{2}x}{\displaystyle dt^{2}}-(\lambda_{1}-\mu-x^{2})\frac{\displaystyle dx}{\displaystyle dt}+x=0,\\
&\frac{\displaystyle d^{2}y}{\displaystyle dt^{2}}-(\lambda_{2}-\mu-y^{2})\frac{\displaystyle dy}{\displaystyle dt}+(1+\delta)y=\mu\frac{\displaystyle dx}{\displaystyle dt}.\\
\end{split}
\end{equation}

The chart of dynamic regimes for the system (2) is given on fig.~6 for $\lambda_{1}=2$, $\lambda_{2}=1$. Comparing this chart with the chart for the case of mutual influence of oscillators (see fig.~4) one can see that formation of the broadband synchronization strip is the result of the influence of the first oscillator that is behind the threshold of the Andronov -- Hopf bifurcation upon the second oscillator, oscillations of which would damp without such an influence. Size and structure of the broadband synchronization area on fig.~4 and fig.~6 are similar. The only difference is that without the mutual influence the oscillation death is observed at any point of parameter space satisfying $\lambda_{1}<\mu$ and $\lambda_{2}<\mu$. In case of absence of the second oscillator's influence upon the first one this fact has simple explanation: the first oscillator appears to be before the threshold of the Andronov -- Hopf bifurcation taking place when $\lambda_{1}-\mu=0$ and its oscillations are damping. The second oscillator is also below the threshold of the Andronov -- Hopf bifurcation, because $\mu>\lambda_{1}>\lambda_{2}$, and its oscillations are also damping without the excitation of the first oscillator.

However, our consideration is valid only when $\lambda_{1}$ and $\lambda_{2}$ values are essentially different. If these parameters values are close, this argumentation can't be used.

\section{\normalsize{The analysis of the broadband synchronization by means of the abridged equations}}

\hspace*{\parindent}It is interesting to find out, which elements of outlined picture one can reveal within the framework of quasiharmonic approximation. We define for it
\begin{equation}
x=\frac{\displaystyle 1}{\displaystyle 2}(a e^{it}+a^{*} e^{-it}),\quad y=\frac{\displaystyle 1}{\displaystyle 2}(b e^{it}+b^{*} e^{-it}),
\end{equation}

and use the traditional additional assumption of
\begin{equation}
\frac{\displaystyle 1}{\displaystyle 2}(\dot{a} e^{it}+\dot{a}^{*} e^{-it})=0,\quad \frac{\displaystyle 1}{\displaystyle 2}(\dot{b} e^{it}+\dot{b}^{*} e^{-it})=0.
\end{equation}

After the averaging we obtain equations for the complex amplitudes
\begin{equation}
\begin{split}
&\frac{\displaystyle da}{\displaystyle dt}=\frac{\displaystyle \lambda_{1} a}{\displaystyle 2}-\frac{\displaystyle |a|^{2}a}{\displaystyle 8}+\frac{\displaystyle 3i\beta|a|^{2}a}{\displaystyle 8}-\frac{\displaystyle \mu}{\displaystyle 2}(a-b),\\
&\frac{\displaystyle db}{\displaystyle dt}=\frac{\displaystyle \lambda_{2} b}{\displaystyle 2}-\frac{\displaystyle |b|^{2}b}{\displaystyle 8}+\frac{\displaystyle 3i\beta|b|^{2}b}{\displaystyle 8}-\frac{\displaystyle \mu}{\displaystyle 2}(b-a)+\frac{\displaystyle i\delta b}{\displaystyle 2}.\\
\end{split}
\end{equation}
After the transformation of variables: $\tau=t/2$, $z=a/2$, $\omega=b/2$ the equations (5) yield
\begin{equation}
\begin{split}
&\frac{\displaystyle dz}{\displaystyle dt}=\lambda_{1}z-|z|^{2}z+3i\beta|z|^{2}z+\mu(\omega-z),\\
&\frac{\displaystyle d\omega}{\displaystyle dt}=\lambda_{2}\omega-|\omega|^{2}\omega+3i\beta|\omega|^{2}\omega+\mu(z-\omega)+i\delta\omega.\\
\end{split}
\end{equation}

Setting $z(t)=R(t)\exp(i\varphi_{1})$ and $\omega(t)=r(t)\exp(i\varphi_{2})$ we obtain equations for the amplitudes $R$, $r$ and their phases $\varphi_{1}$ and $\varphi_{2}$
\begin{equation}
\begin{split}
&\frac{\displaystyle dR}{\displaystyle d\tau}=R(\lambda_{1}-\mu)-R^{3}+\mu r \cos(\varphi_{2}-\varphi_{1}),\\
&\frac{\displaystyle dr}{\displaystyle d\tau}=r(\lambda_{2}-\mu)-r^{3}+\mu R \cos(\varphi_{1}-\varphi_{2}),\\
&\frac{\displaystyle d\varphi_{1}}{\displaystyle d\tau}=3\beta R^{2}+\frac{\displaystyle r}{\displaystyle R}\mu \sin(\varphi_{2}-\varphi_{1}),\\
&\frac{\displaystyle d\varphi_{2}}{\displaystyle d\tau}=3\beta r^{2}+\frac{\displaystyle R}{\displaystyle r}\mu \sin(\varphi_{1}-\varphi_{2})+\delta.\\
\end{split}
\end{equation}

There is only the phase difference in the equations for amplitudes. Therefore, one can subtract the second equation for the phase from the first one and obtain the following abridged equations containing the relative phase of oscillators $\psi=\varphi_{1}-\varphi_{2}$:
\begin{equation}
\begin{split}
&\frac{\displaystyle dR}{\displaystyle d\tau}=R(\lambda_{1}-\mu)-R^{3}+\mu r \cos \psi,\\
&\frac{\displaystyle dr}{\displaystyle d\tau}=r(\lambda_{2}-\mu)-r^{3}+\mu R \cos \psi,\\
&\frac{\displaystyle d\psi}{\displaystyle d\tau}=-\delta+3\beta (R^{2}-r^{2})-\mu(\frac{\displaystyle r}{\displaystyle R}+\frac{\displaystyle R}{\displaystyle r})\sin \psi.\\
\end{split}
\end{equation}

We have obtained the system of equations similar to [1] but it is taking the possible non-identity of controlling parameters into account.

Evaluate the lower boundary of the broadband synchronization area analytically using the abridged equations (8). Consider the stationary system of (8):
\begin{equation}
\begin{split}
&0=R(\lambda_{1}-\mu)-R^{3}+\mu r \cos \psi,\\
&0=r(\lambda_{2}-\mu)-r^{3}+\mu R \cos \psi,\\
&\delta=3\beta (R^{2}-r^{2})-\mu(\frac{\displaystyle r}{\displaystyle R}+\frac{\displaystyle R}{\displaystyle r})\sin \psi.\\
\end{split}
\end{equation}

We suppose that the summands $\mu R \cos \psi$ and $\mu r \cos \psi$ are small on the boundary of the synchronization area. Then the evaluations of the steady amplitudes of limit cycles are
\begin{equation}
R\approx\sqrt{\lambda_{1}-\mu},\quad r\approx\sqrt{\lambda_{2}-\mu}.
\end{equation}

Since radical expressions should be non-negative it is necessary to satisfy the conditions $(\lambda_{1}\geq\mu)$ and $(\lambda_{2}\geq\mu)$. Substituting (10) into the third equation of (9), we obtain
\begin{equation}
\delta=3\beta((\sqrt{\lambda_{1}-\mu})^{2}-(\sqrt{\lambda_{2}-\mu})^{2})-\mu \left(\frac{\displaystyle \sqrt{\lambda_{2}-\mu}}{\displaystyle \sqrt{\lambda_{1}-\mu}}+\frac{\displaystyle \sqrt{\lambda_{1}-\mu}}{\displaystyle \sqrt{\lambda_{2}-\mu}}\right)\sin \psi.
\end{equation}

This equation is a sort of the well-known stationary Adler equation [1]. Its stability threshold is defined by the condition of $\sin\psi=1$. But then we have $\cos\psi=0$ and our assumption of the neglected terms smallness in the first and the second equations (9) is justified.

Consider the case of absence of the phase nonlinearity. Then (11) yields
\begin{equation}
\delta=\pm\mu\left(\frac{\displaystyle \sqrt{\lambda_{2}-\mu}}{\displaystyle \sqrt{\lambda_{1}-\mu}}+\frac{\displaystyle \sqrt{\lambda_{1}-\mu}}{\displaystyle \sqrt{\lambda_{2}-\mu}}\right).
\end{equation}
$\delta(\mu)$-function given by the relation (12) defines the boundary of the synchronization area which ought to move to the lower boundary of the infinite synchronization band according to fig.~3 and fig.~4. Denote this branch as $\delta_{+}(\mu)$ and consider its asymptotic properties in case of $\lambda_{1}>\lambda_{2}$. If the coupling parameter tends to the value of controlling parameter of the second oscillator, then $(\lambda_{2}-\mu)\rightarrow0$. It means that $\sqrt{(\lambda_{2}-\mu)/(\lambda_{1}-\mu)}\ll\sqrt{(\lambda_{1}-\mu)/(\lambda_{2}-\mu)}$ and $\delta(\mu)$-function tends to the infinity, i.e. $\lim_{\mu \to \lambda_{2}}\delta_{+}(\mu)=\infty$. So it must be $\mu\rightarrow\lambda_{2}$ when $\delta\rightarrow\infty$ for the graph of inverse function $\mu=\mu(\delta)$. It means that the boundary of the synchronization area extends arbitrarily far to the area of large frequency mismatches.

$\delta(\mu)$-function given by the relation (12) defines the lower boundary of the infinitely long synchronization band. The upper boundary of this band is the boundary of the oscillation death area when the trivial rest state of point of origin becomes stable. Use the abridged equations obtained before for the system of coupled van der Pol - Duffing oscillators (6) for the analytical estimation of this boundary. After the linearization of the system (6) about a point of origin we obtain
\begin{equation}
\begin{split}
&\frac{\displaystyle dz}{\displaystyle d\tau}=\lambda_{1}z+\mu(\omega-z),\\
&\frac{\displaystyle d\omega}{\displaystyle d\tau}=\lambda_{2}\omega+\mu(z-\omega)+i\delta\omega.
\end{split}
\end{equation}
For the investigation of the stability of zero rest state of the obtained linear system set
\begin{equation}
z\sim e^{(\eta+i\varepsilon)t},\quad \omega\sim e^{(\eta+i\varepsilon)t},
\end{equation}
where $\eta,\varepsilon$ are real numbers. It is necessary to satisfy the condition $\eta<0$ for the realization of the oscillation death regime. Correspondingly, there is $\eta=0$ on the boundary of the oscillation death area. Then
\begin{equation}
z\sim e^{i\varepsilon t},\quad \omega\sim e^{i\varepsilon t}.
\end{equation}
Substitution of (15) in the system (13) yields the relation
\begin{equation}
\begin{split}
&(i\varepsilon-\lambda_{1}+\mu)(i\varepsilon-\lambda_{2}+\mu-i\delta)=\mu^{2}\Leftrightarrow\\
\Leftrightarrow \varepsilon^{2}+\varepsilon(-\delta+&i(\lambda_{1}+\lambda_{2}-2\mu))+(\mu(\lambda_{1}+\lambda_{2})-\lambda_{1}\lambda_{2}+i\delta(\mu-\lambda_{1}))=0.
\end{split}
\end{equation}
Denote
\begin{equation}
\begin{split}
&\alpha_{x}=-\delta,\\
&\alpha_{y}=\lambda_{1}+\lambda_{2}-2\mu,\\
&\beta_{x}=\mu(\lambda_{1}+\lambda_{2})-\lambda_{1}\lambda_{2},\\
&\beta_{y}=\delta(\mu-\lambda_{1}).
\end{split}
\end{equation}
Then
\begin{equation}
\varepsilon^{2}+\varepsilon(\alpha_{x}+i\alpha_{y})+(\beta_{x}+i\beta_{y})=0,\\
\end{equation}
where $\alpha_{x},\alpha_{y},\beta_{x},\beta_{y}$ are real numbers. Therefore one can divide the equation (18) into the real and imaginary parts: $\varepsilon^{2}+\varepsilon\alpha_{x}+\beta_{x}=0$ and $\varepsilon\alpha_{y}+\beta_{y}=0$. It follows that
\begin{equation}
\varepsilon=-\frac{\displaystyle \beta_{y}}{\displaystyle \alpha_{y}},\quad \frac{\displaystyle \beta_{y}^{2}}{\displaystyle \alpha_{y}^{2}}-\frac{\displaystyle \beta_{y}\alpha_{x}}{\displaystyle \alpha_{y}}+\beta_{x}=0.
\end{equation}
Substituting (17) in (19), we obtain after some transformations the relation that defines the boundary of the oscillation death area and the upper boundary of the synchronization band, correspondingly
\begin{equation}
\delta^{2}=\frac{\displaystyle (\mu(\lambda_{1}+\lambda_{2})-\lambda_{1}\lambda_{2})(\lambda_{1}+\lambda_{2}-2\mu)^{2}}{\displaystyle (\lambda_{1}-\mu)(\lambda_{2}-\mu)}.
\end{equation}
In special case when $\lambda_{1}=\lambda_{2}=\lambda$ the relation (20) yields
\begin{equation}
\delta^{2}=4\lambda(2\mu-\lambda),
\end{equation}
and the oscillation death area is defined by the certain [1, 2] two-sided inequality
\begin{equation}
\lambda<\mu<\frac{\displaystyle 1}{\displaystyle 2}\left(\lambda+\frac{\displaystyle \delta^{2}}{\displaystyle 4\lambda}\right),
\end{equation}

Investigation of the behavior of $\delta(\mu)$-function given by the relation (20) yields its asymptotic features
\begin{equation}
\lim_{\mu\to\lambda_{1}}\delta(\mu)=\infty,
\end{equation}
\begin{equation}
\lim_{\mu\to\infty}\delta(\mu)=2\sqrt{\mu(\lambda_{1}+\lambda_{2})}.
\end{equation}

The graphs of $\delta(\mu)$-functions given by the relations (12) and (20) are shown in fig.~7. Inside the grey area the rest state of the Adler equation exists and, correspondingly, the synchronization of oscillators is possible. We see that there is a synchronization band, which is infinitely long in eigenfrequency mismatch, in the coupling parameter domain of $\lambda_{2}<\mu<\lambda_{1}$. Consequently, the analysis that was made above allows to show the presence of the infinite on frequency mismatch synchronization band. It should be noticed that this analysis does not allow to diagnosticate the multiple synchronization regimes and, therefore, it is a question of the synchronization of 1/1-type.

\section{\normalsize{The broadband synchronization in system with the phase nonlinearity}}

\hspace*{\parindent}We consider now the case when there is the phase nonlinearity presented in the system. The question of interest is how the phase nonlinearity influences the broadband synchronization domain and the overall structure of the parameter plane (eigenfrequency mismatch - coupling strength). One can determine the lower boundary of the synchronization band from the equation (11) using the fact that this equation is the stationary Adler equation. This boundary is given by the relation
\begin{equation}
\delta=\pm\mu \left(\frac{\displaystyle \sqrt{\lambda_{2}-\mu}}{\displaystyle \sqrt{\lambda_{1}-\mu}}+\frac{\displaystyle \sqrt{\lambda_{1}-\mu}}{\displaystyle \sqrt{\lambda_{2}-\mu}}\right)+3\beta((\sqrt{\lambda_{1}-\mu})^{2}-(\sqrt{\lambda_{2}-\mu})^{2}).
\end{equation}

The graph of the $\delta(\mu)$-function given by the relation (25) is shown in fig.~7 by the dashed line. At the same time, the upper boundary of the synchronization band shown in fig.~7 is determined, as before, by the relation (20) since it corresponds to the linearized system and is valid independently of the presence or absence of the phase nonlinearity. One can see that presence of the phase nonlinearity leads to the displacement of the synchronization tongue along the axis of eigenfrequency mismatch and does not influence on the broadband synchronization area in the case of large mismatches.

Dynamic regime charts obtained via numerical simulations of the system of coupled van der Pol -- Duffing oscillators are shown in fig. 8 for different values of $\lambda_{1}$ and $\lambda_{2}$ in presence of the phase nonlinearity in system. One can see that the synchronization tongue shifts to the right along the axes of eigenfrequency mismatch with the increase of the parameter of nonlinearity $\beta$, just as the analysis of the Adler equation predicts. One can see though that the phase nonlinearity influences strongly on the areas with fractional rotation numbers. Inside the synchronization band they decrease slightly in size, change their own shape and their frequency threshold shifts strongly to the area of large mismatches (to the right).

Phase plane portraits for different points of the parameter plane are given in fig.~8~(b) in the $(x,\dot{x})$-plane (left one) and in the $(y,\dot{y})$-plane (right one). Moving along the synchronization band, one can see the transition from the synchronization regime of 1/1-type to the regime of 1/3-type. One can note one more feature of the phase plane portraits in the area of 1/3 in fig.~8 in comparison with the fig.~3. It is the fact that "loops" of the second oscillator's attractor on the phase plane portrait may enclose a coordinate origin.

\section{\normalsize{System with non-identity in value of nonlinear dissipation}}

\hspace*{\parindent}In described picture of interaction between oscillators the first oscillator dominates more over the second one with increase of the frequency mismatch in the broadband synchronization area, when $\lambda_{2}<\mu<\lambda_{1}$. It becomes apparent in noticeable downsizing of the second oscillator's limit cycle in relation to the size of the first oscillator's limit cycle. (Pay attention to the scales in coordinate axis on the phase plane portraits in fig.~3). But it is possible to obtain synchronization regimes when oscillators are equivalent also in case of large frequency mismatch. Consider for that the following system of equations:
\begin{equation}
\begin{split}
&\frac{\displaystyle d^{2}x}{\displaystyle dt^{2}}-(\lambda_{1}-x^{2})\frac{\displaystyle dx}{\displaystyle dt}+x+\beta x^{3}+\mu(\frac{\displaystyle dx}{\displaystyle dt}-\frac{\displaystyle dy}{\displaystyle\displaystyle dt})=0,\\
&\frac{\displaystyle d^{2}y}{\displaystyle dt^{2}}-(\lambda_{2}-\gamma y^{2})\frac{\displaystyle dy}{\displaystyle dt}+(1+\delta)y+\beta y^{3}+\mu(\frac{\displaystyle dy}{\displaystyle dt}-\frac{\displaystyle dx}{\displaystyle dt})=0.\\
\end{split}
\end{equation}
Here additional parameter $\gamma$ characterizes nonlinear dissipation of the second oscillator. This parameter together with the parameter $\lambda$ defines the size of the limit cycle in autonomous behavior. Using quasiharmonic approximation one can estimate corresponding sizes of the limit cycles: for the first oscillator it is $\sqrt{\lambda_{1}}$ and for the second one it is $\sqrt{\lambda_{2}/\gamma}$. So using small values of parameter $\gamma$ one can greatly enlarge the size of the second oscillator's limit cycle.

Dynamic regime chart for the system (26) for $\lambda_{1}=2, \lambda_{2}=1, \gamma=0.01, \beta=0$ and several typical phase plane portraits are shown in fig.~9. One can see that the broadband synchronization remains in that case. Moreover, the main synchronization tongue appears to have an extension within which the second oscillator dominates over the first one (see fig.~9~(a)). In areas of multiple synchronization it may be both the situation of equivalent oscillators and the situation of changing of "leading" oscillator, when the second oscillator dominates over the second one (see fig.~9~(b)). "Equality" or oscillator's domination may be determined from the shape of the phase plane portrait and the size of the attractor.

\section{\normalsize{Conclusions}}

\hspace*{\parindent}Dissipatively coupled non-identical in controlling parameters van der Pol oscillators demonstrate interesting synchronization features. On the parameter plane (frequency mismatch -- coupling value) the boundary between the oscillation death area and the area of quasiperiodic regimes turns from the line into the band of finite width which stretches along the mismatch parameter axis into the area of large mismatch values. In the case of small non-identity the synchronization with the rotation number of 1/1, generally, corresponds to this band. However, there are also regimes of 1/3, 1/5 types, etc. and their domains of stability on the parameter plane look like "degraded" tops of corresponding synchronization tongues. Areas corresponding to these regimes expand essentially with the increase of non-identity, filling, practically, the whole band. The oscillation regime when one oscillator dominates distinctly over another oscillator is realized in that case. Areas with fractional rotation numbers decrease in size and their frequency threshold moves towards the large mismatch values in the case of presence of the phase nonlinearity of Duffing oscillator type. The size of the "broadband" synchronization area increases with insertion of a non-identity in value of nonlinear dissipation due to the formation of "outgrowth" from the main synchronization tongue. And at the same time regimes, for which oscillators are equivalent and the changing of "leading" oscillator may take place, appear.

\section*{\normalsize{Acknowledgements}}
\hspace*{\parindent}This work was supported by RFBR (Project No. 06-02-16773) and foundation of nonprofit programs "Dynasty".

\renewcommand{\refname}{\normalsize{References}}

\renewcommand\figurename{Fig.}
\newpage
\begin{figure}[!h]
\begin{center}
\includegraphics[scale=1]{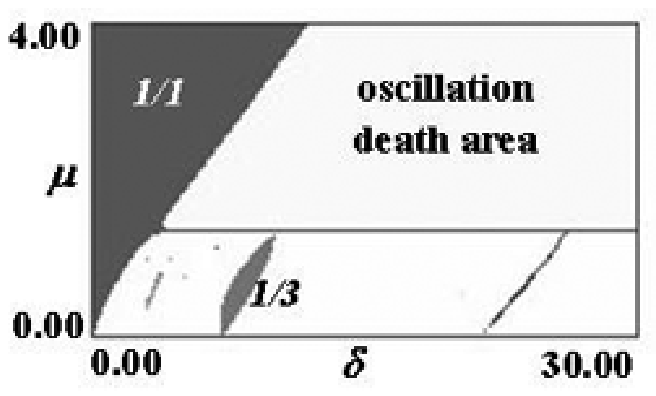}
\end{center}
\raggedright
\caption{Dynamic regime chart for the system (1) for $\lambda_{1}=\lambda_{2}=1$, $\beta=0$.}
\end{figure}

\newpage
\begin{figure}[!h]
\begin{center}
\includegraphics[scale=1]{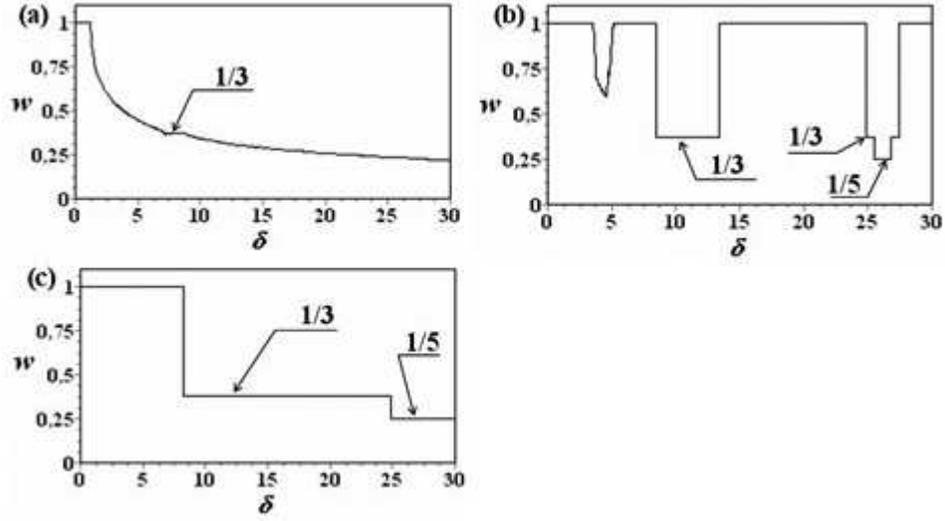}
\end{center}
\raggedright
\caption{Dependence of the rotation number $w$ upon the frequency mismatch $\delta$ for \mbox{(a)~$\lambda_{1}=1.25$}, $\lambda_{2}=1$, $\mu=0.5$;(b)~$\lambda_{1}=1.25$, $\lambda_{2}=1$, $\mu=1.05$; (c)~$\lambda_{1}=2$, $\lambda_{2}=1$, $\mu=1.5$.}
\end{figure}

\newpage
\begin{figure}[!h]
\begin{center}
\includegraphics[scale=1]{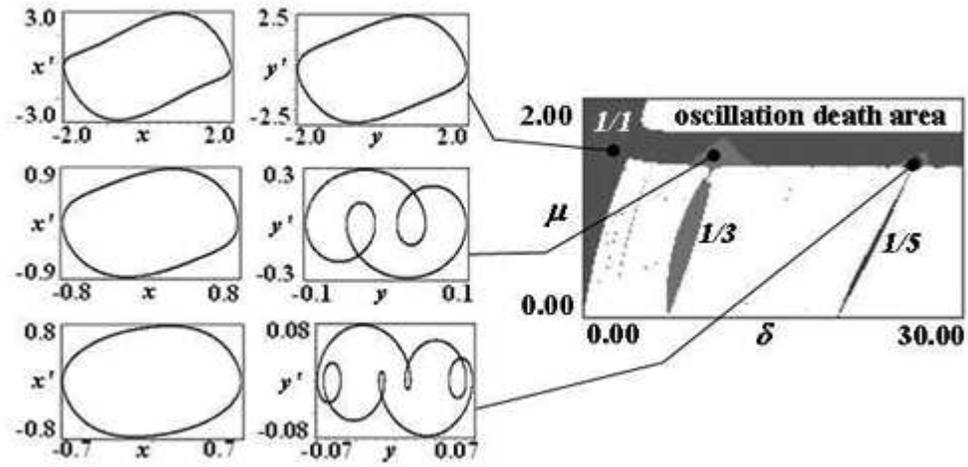}
\end{center}
\raggedright
\caption{Dynamic regime chart for the system (1) for $\lambda_{1}=1.25$, $\lambda_{2}=1$, $\beta=0$ and phase plane portraits in characteristic areas of the parameter plane.}
\end{figure}

\newpage
\begin{figure}[!h]
\begin{center}
\includegraphics[scale=1]{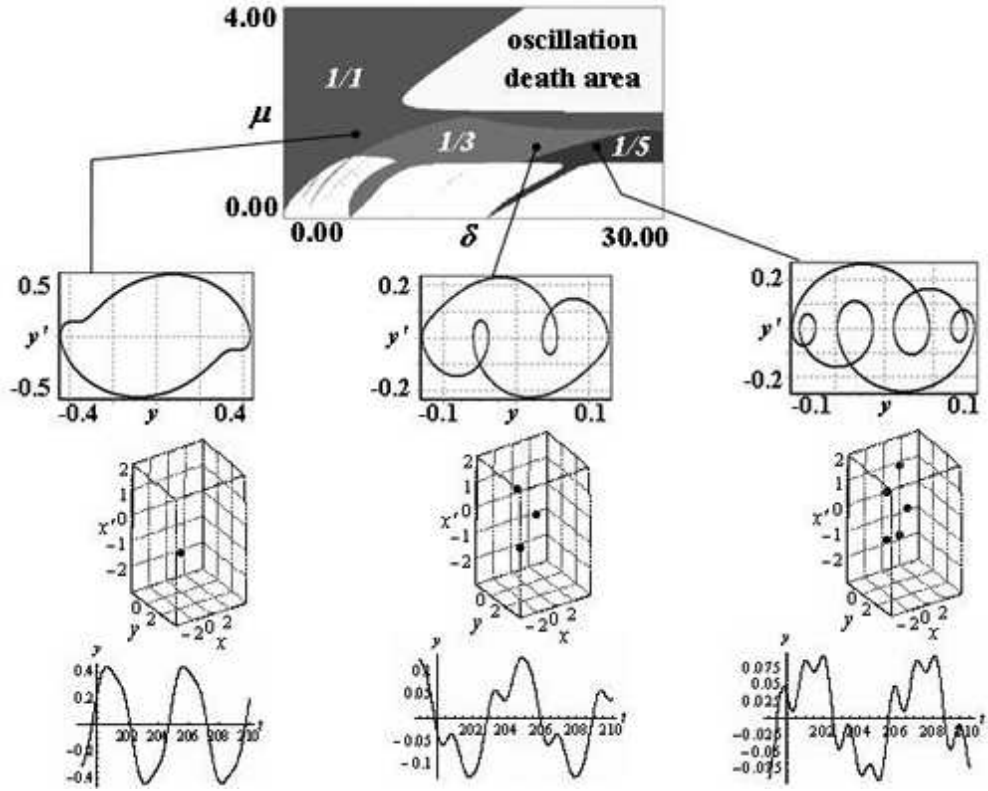}
\end{center}
\raggedright
\caption{Dynamic regime chart for the system (1) for $\lambda_{1}=2$, $\lambda_{2}=1$, $\beta=0$ and phase plane portraits of the second oscillator, its Poincare sections and realizations in typical areas of the parameter plane.}
\end{figure}

\newpage
\begin{figure}[!h]
\begin{center}
\includegraphics[scale=1]{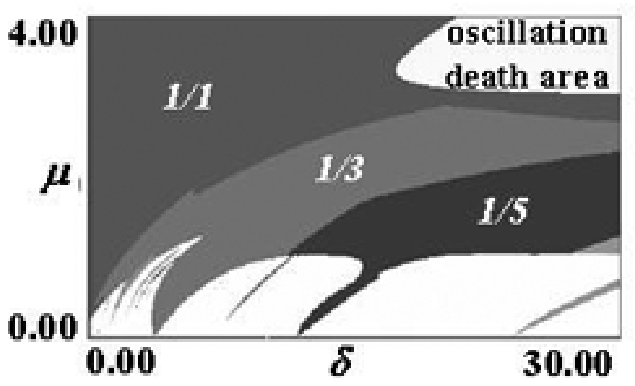}
\end{center}
\raggedright
\caption{Dynamic regime chart for the system (1) for $\lambda_{1}=3$, $\lambda_{2}=1$, $\beta=0$.}
\end{figure}

\newpage
\begin{figure}[!h]
\begin{center}
\includegraphics[scale=1]{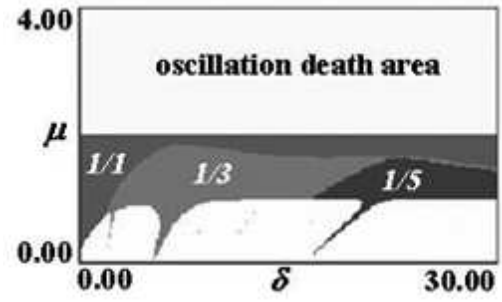}
\end{center}
\raggedright
\caption{Dynamic regime chart for the system (2) with the "shut off" influence of the second oscillator upon the first one for  for $\lambda_{1}=2$, $\lambda_{2}=1$.}
\end{figure}

\newpage
\begin{figure}[!h]
\begin{center}
\includegraphics[scale=1]{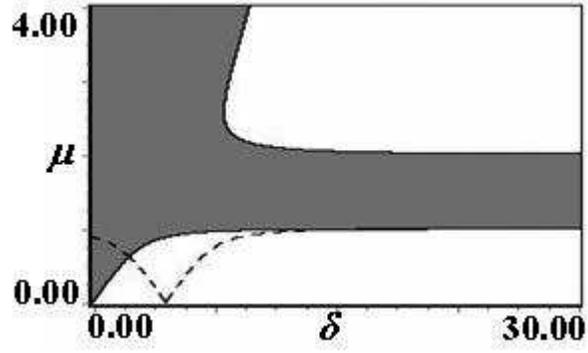}
\end{center}
\raggedright
\caption{Synchronization area of 1/1-type defined analytically within the framework of the phase approximation for $\lambda_{1}=2$, $\lambda_{2}=1$, $\beta=0$  shown by the grey color. Boundary of the synchronization area in presence of the phase nonlinearity $\beta=1$ is shown by the dashed line.}
\end{figure}

\newpage
\begin{figure}[!h]
\begin{center}
\includegraphics[scale=1]{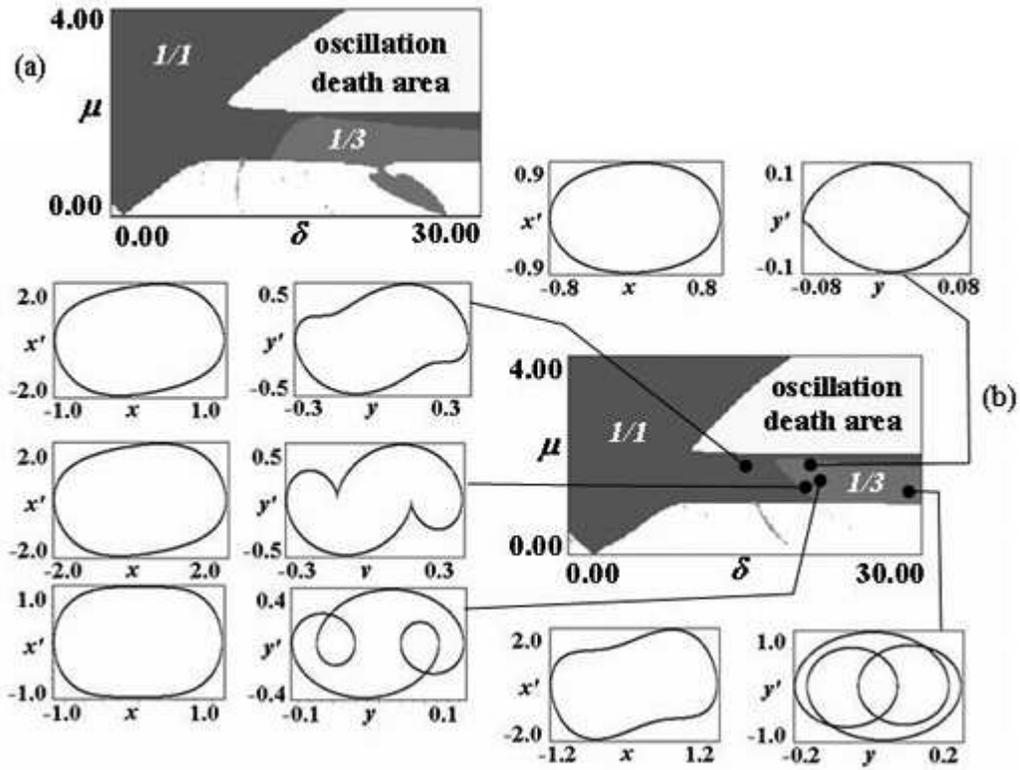}
\end{center}
\raggedright
\caption{Dynamic regime charts for the system (1) for $\lambda_{1}=2$, $\lambda_{2}=1$ and (a)~$\beta=0.5$; (b)~$\beta=1$ with phase plane portraits at characteristic points.}
\end{figure}

\newpage
\begin{figure}[!h]
\begin{center}
\includegraphics[scale=1]{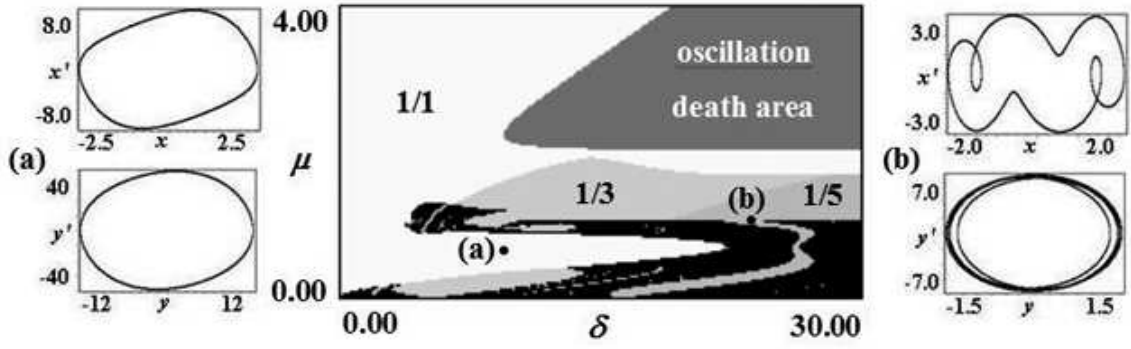}
\end{center}
\raggedright
\caption{Dynamic regime charts for the system (26) for $\lambda_{1}=2$, $\lambda_{2}=1$, $\gamma=0.01$, $\beta=0$ and typical phase plane portraits of the first (on top) and the second (from below) oscillators.}
\end{figure}

\end{document}